# Stable antiferromagnetic graphone

D. W. Boukhvalov

*Computational Materials Science Center, National Institute for Materials Science,*

*1-2-1 Sengen, Tsukuba, Ibaraki 305-0047, Japan*

*Density functional modeling of atomic structure and calculation of electronic structure of one-side one-sublattice functionalized graphene (graphone) are performed for hydrogen and fluorine adatoms. Shown that using of fluorine for functionalization not on enhance stability of compound but also provide switch of magnetic ground state from ferro- to antiferromagnetic. Half-metallic ferromagnetic state in fluorine based graphone is also discussed.*

E-mail: D.Bukhvalov@science.ru.nl

# 1. Introduction

Graphene is the one of most attractive materials for post-silicon electronics [1-3]. Chemical functionalization of graphene [4] is the source for tune electronic properties of graphene based materials. Theoretical prediction [5, 6] and further production [7] of 100% covered by hydrogen graphene (graphane) show us the new frontiers in graphene physics and chemistry. Recent theoretical model [8] of graphene with hydrogenated only one of sublattices from one side (graphone, see Fig. 1a) predict coexistence there room-temperature ferromagnetism and energy gap.

Main principles of graphene functionalization developed in our previous works [4, 6] argue for the instability of one-sublattice functionalization. In Ref. [8] stability of graphone was checked by molecular dynamics methods. Further examination of graphone stability has required calculation of the activation energies of the hydrogen atoms removal and migration barriers for hydrogen fluidity to other carbon sites. Test of other adatoms are suggested by easy graphane total dehydrogenation by annealing at 400C. Our previous calculations [4] and recent experimental works [9-12] have shown that fluorine is the most suitable chemical space for enhancement stability of graphone.

# 2. Computational models and method

Density functional theory (DFT) calculations had carried out with the same pseudopotential code SIESTA [13] used for our previous modeling of functionalized graphene [4, 6, 14]. All calculations are done in the generalized gradient approximation (GGA-PBE) [15] which is more suitable for description of graphene-adatom chemical bonds [6]. For the modeling of graphone I have used supercell contained 50 carbon atoms

(see Fig. 1) within periodic boundary conditions. The distance between equivalent atoms in supercells is 5 lattice parameters (about 1.2 nm). Thus size of supercell is guaranteed absence of any overlap between distorted areas around chemisorbed along the ribbon groups [4].

All calculations were carried out for energy mesh cut off 360 Ry and k-point mesh 8×8×2 in Mokhorst-Park scheme [16]. During the optimization, the electronic ground state was found self-consistently using norm-conserving pseudo-potentials for cores and a double-$\zeta$ plus polarization basis of localized orbitals for carbon and oxygen, and double-$\zeta$ basis for hydrogen. Optimization of the bond lengths and total energies was performed with an accuracy of 0.04 eV/Å and 1 meV, respectively. When drawing the pictures of density of states, a smearing of 0.05 eV was used.

For examine activation energies I have performed differences of total energies pristine graphone (Fig. 1a) and graphone with one and two adatoms shifted far from its plane (Figs. 1b and c respectively). This method was used for calculation of activation energies of similar systems [14, 17]. The migration processes are characterizing by two numbers. First is the total energy difference between initial system (pristine graphone Fig. 1a) and final system with one adatom moved for carbon of another sublattice (Fig. 1d). Second number is the energy barrier was calculated with using methods proposed for migration of adatoms on graphene [18-20].

## 3. Results and discussions

### 3.1. Stability of hydrogenated graphone

For check the correctness of used method, model and pseudopotential choice results from Ref. [8] were reproduced. Optimized atomic structure (Fig. 1a and 2), electronic structure (Fig. 3) and energy difference between ferromagnetic and antiferromagnetic configuration were obtained in rather good agreement to the previous results. For check the stability calculation of atomic structure and the total energy of graphone with pushed far from the flat one and two hydrogen atoms (see Fig. 1b and c) were performed. Results of calculations reported in Table 1 argue that activation barrier for start of dehydrogenation process (energy required for remove first hydrogen atom) is high enough and comparable to graphane [14]. This stability to dehydrogenation is in agreement with previous MD simulation for graphane [8]. How suggest us previous graphene functionalization studies [4, 6] the magnetic state could be destroyed by adatom reconstruction due to migration. For describe the first state of these processes calculation of migration energy and barrier for one hydrogen adatom migration was performed. In contrast with removal of hydrogen the migration of adatom is very energetically favorable and the migration barrier for this move very small. That makes hydrogen based graphone rather unstable for using in devices.

### 3.2. Structural and magnetic properties of fluorinated graphone

Recent calculations [4] and experiments [9-12] suggest us that 100% fluorinated graphene is more stable than graphane. The results of atomic structure optimization presented on Fig. 2. Lattice distortions between hydrogenated and fluorinated graphones is not significant but changes of electronic structure (Fig. 3) could be described as dramatic. In the case of hydrogenated graphone observed two very well separated sharp picks near Fermi level corresponding with unpaired electrons in non-functionalized carbon sublattice. When we substitute hydrogen ($1s^1$) to fluorine ($2s^2\ 2p^5$) the strong hybridization between carbon and fluorine *2p* orbitals are observed. These carbon-fluorine interactions provide smearing of localized bands corresponding to unpaired electrons. For the case of ferromagnetic configuration (Fig. 4a) huge graphene sheet corrugation provides formation of midgap states [19, 21-23] and near-midgap states in fluorinated graphone. The formation of midgap states is very energetically unfavorable and in nonmagnetic graphenic systems could be solved by further chemical functionalization [23]. In magnetic graphone realize another route for remove the midgap states. Similarly to graphene on hexagonal boron nitride [24] division of one sublattice provides opening the energy gap. In magnetic systems this separation could be realized over transition to antiferromagnetic configurations. Both antiferromagnetic configurations (Fig. 4b, c) are more stable than ferromagnetic. Turn from ferromagnetic to antiferromagnetic configuration provide the energy gap opening. Anther effect of strong carbon-fluorine interaction is the increasing about three times the total energy difference between ferromagnetic and antiferromagnetic configurations (Fig 4a and b).

Large binding energy between graphene and fluorine provide increasing the activation energies for removal of one and two adatoms from the fluorinated graphone (see Table 1). The expansion of carbons lattice is increase migration distance and stronger corrugation of graphene flat make the final state of migration of fluorine also unstable. It makes the migrations of adatoms of fluorinated graphone energetically unfavorable.

## 4. Conclusions

Performed DFT modelings have shown that fluorinated graphone is rather stable that hydrogenated one. It makes fluorinated graphone more suitable for the further applications in electronic and spintronic devices. The interactions between carbon and fluorine *2p* orbitals dramatically change electronic structure and magnetic properties of fluorinated graphone. In contrast with hydrogen functionalized graphone fluorinated compound have antiferromagnetic ground state. The switch from antiferromagnetic insulating to ferromagnetic half-metallic state makes fluorinated graphone prospective material for further electronic applications [25].

**Table 1.** Activation and migration energies and migration barriers (both in eV) for graphones functionalized with hydrogen and fluorine.

| Adatom | Remove of *one* adatom | Remove of *two* adatoms | Migration energy | Migration barrier |
|---|---|---|---|---|
| Hydrogen | +1.15 | -5.31 | -1.44 | +0.06 |
| Fluorine | +1.43 | +0.50 | +0.17 | +0.72 |

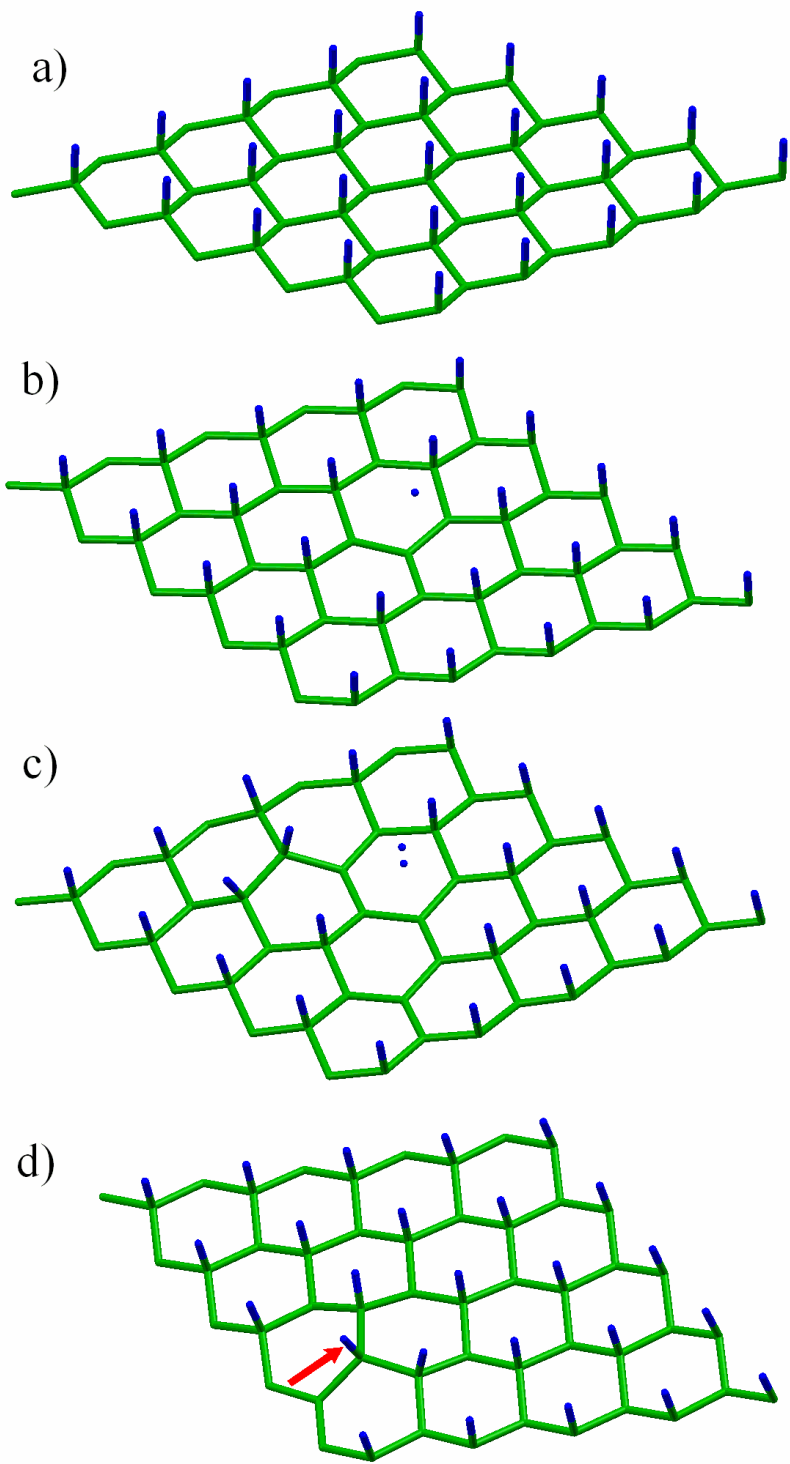

**Fig. 1** Optimized atomic structure of pristine graphone (a), graphone with removed one (b) and two (c) adatoms; final step of the migration of adatom (shown by arrow) from one sublattice to another (d). Carbon atoms are shown by green, hydrogen by blue.

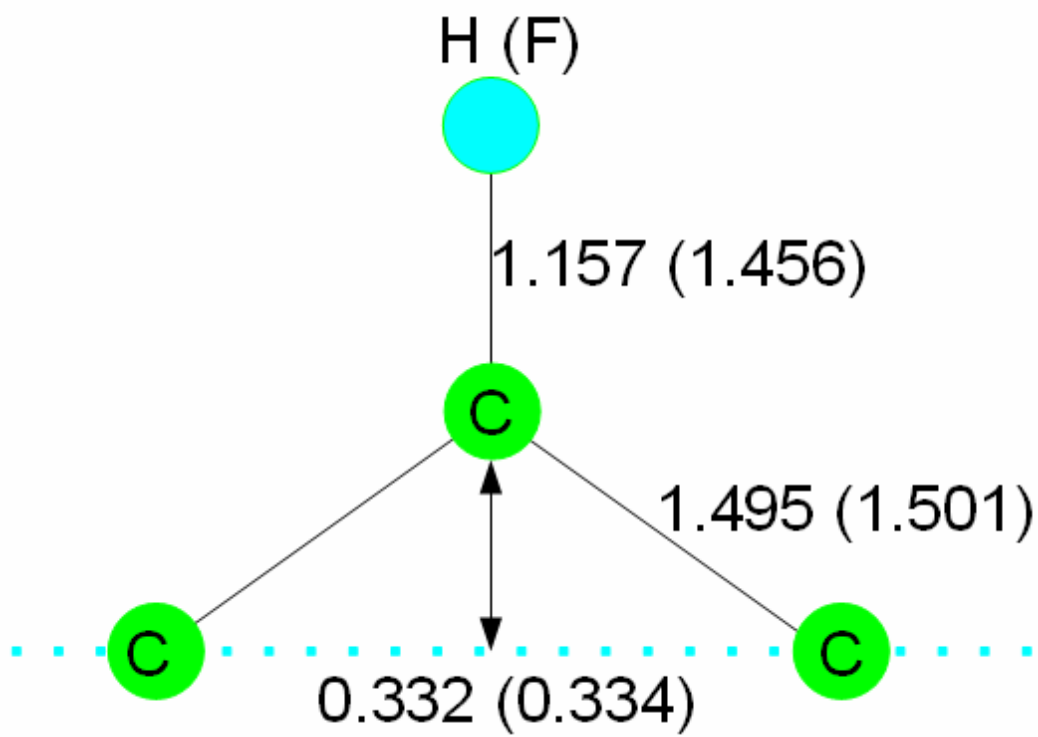

**Fig. 2** A sketch of local atomic structure for graphone with reciprocal interatomic distances for hydrogen and fluorine (in parenthesis) cases. All distances are shown in Å.

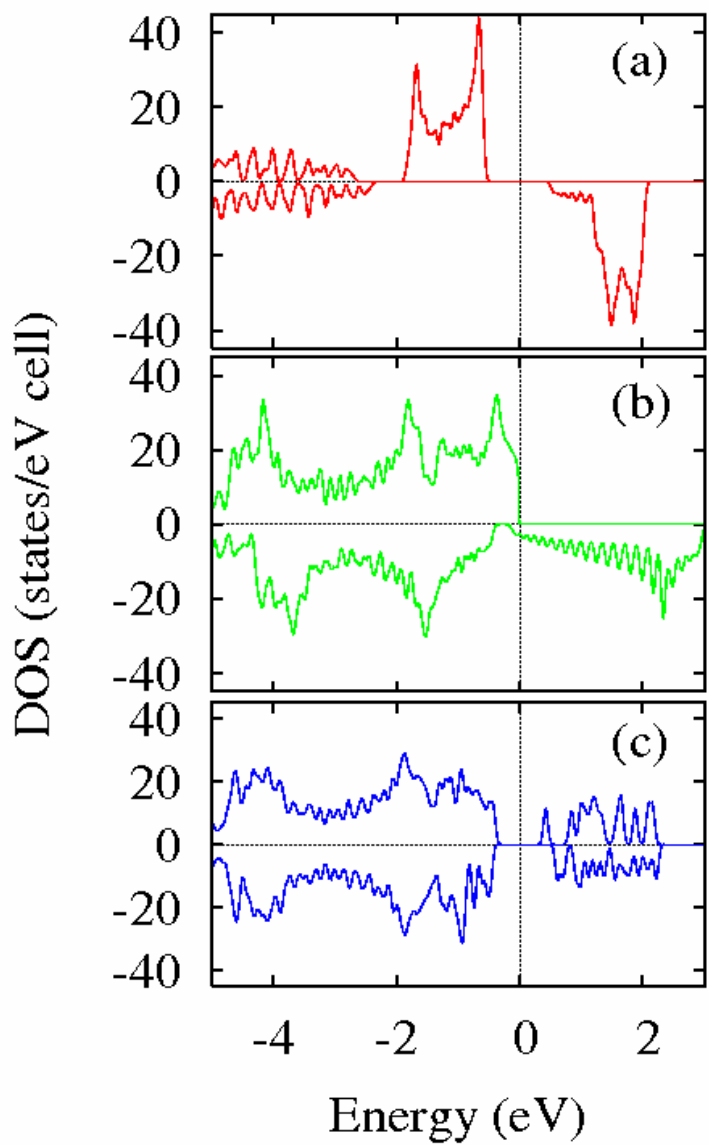

**Fig. 3** Total densities of states for graphones functionalized by hydrogen in ferromagnetic ground state (a) and fluorine in ferromagnetic (b) and antiferromanetic (Fig. 4b) states.

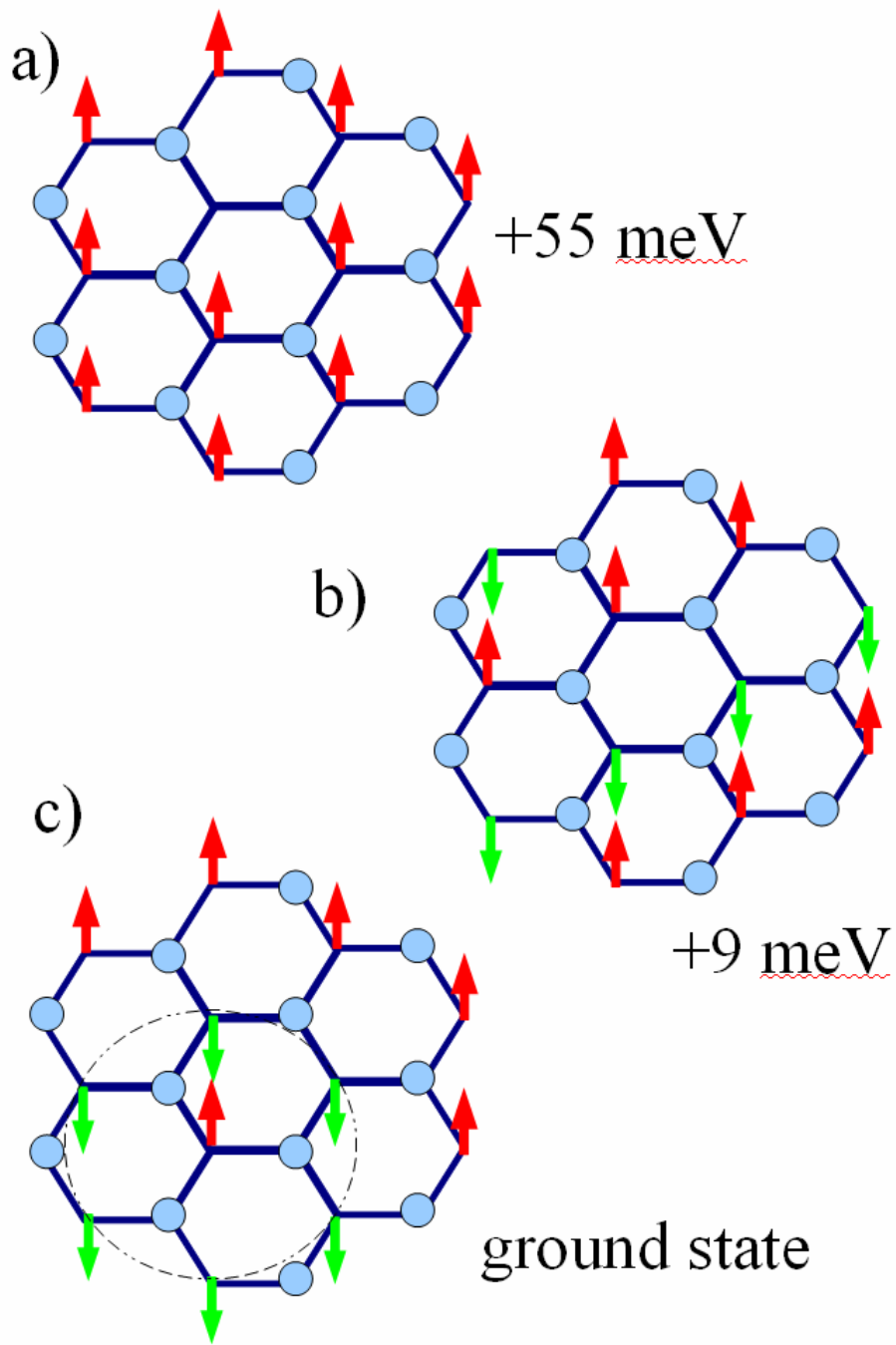

**Fig. 4** Schemes of ferromagnetic (a) and two types of antiferromagnetic orders (b,c) in fluorine functionalized graphone. The numbers are corresponding to relative energies to magnetic ground state per atom.